\documentclass[useAMS,usenatbib,preprint2]{aastex}
\usepackage{amsmath}
\usepackage{graphicx}
\usepackage{tikz}

\shorttitle{BLACK HOLE SPIN IN OJ~287}
\shortauthors{Valtonen et al.}

\begin{document}

\title{PRIMARY BLACK HOLE SPIN IN OJ~287 AS DETERMINED BY THE GENERAL RELATIVITY CENTENARY FLARE}

\author{M.~J.~Valtonen\altaffilmark{1,2}, 
S.~Zola\altaffilmark{3,4}, 
S.~Ciprini\altaffilmark{5,6}, 
A.~Gopakumar\altaffilmark{7}, 
K.~Matsumoto\altaffilmark{8}, 
K.~Sadakane\altaffilmark{8}, 
M.~Kidger\altaffilmark{9}, 
K.~Gazeas\altaffilmark{10}, 
K.~Nilsson\altaffilmark{1}, 
A.~Berdyugin\altaffilmark{2}, 
V.~Piirola\altaffilmark{1,2}, 
H.~Jermak\altaffilmark{11}, 
K.~S.~Baliyan\altaffilmark{12}, 
F.~Alicavus\altaffilmark{13}, 
D.~Boyd\altaffilmark{14}, 
M.~Campas Torrent\altaffilmark{15}, 
F.~Campos\altaffilmark{16}, 
J.~Carrillo G{\'o}mez\altaffilmark{17}, 
D.~B.~Caton\altaffilmark{18}, 
V.~Chavushyan\altaffilmark{19}, 
J.~Dalessio\altaffilmark{20}, 
B.~Debski\altaffilmark{3}, 
D.~Dimitrov\altaffilmark{21},
M.~Drozdz\altaffilmark{4}, 
H.~Er\altaffilmark{22}, 
A.~Erdem\altaffilmark{13}, 
A.~Escartin P{\'e}rez\altaffilmark{23}, 
V.~Fallah Ramazani\altaffilmark{2}, 
A.~V.~Filippenko\altaffilmark{24},
S.~Ganesh\altaffilmark{12}, 
F.~Garcia\altaffilmark{25}, 
F.~G{\'o}mez Pinilla\altaffilmark{26}, 
M.~Gopinathan\altaffilmark{27}, 
J.~B.~Haislip\altaffilmark{28}, 
R.~Hudec\altaffilmark{29}, 
G.~Hurst\altaffilmark{30}, 
K.~M.~Ivarsen\altaffilmark{28},
M.~Jelinek\altaffilmark{29}, 
A.~Joshi\altaffilmark{27}, 
M.~Kagitani\altaffilmark{31},
N.~Kaur\altaffilmark{12}, 
W.~C.~Keel\altaffilmark{32}, 
%P.~Kushwaha\altaffilmark{34}, 
A.~P.~LaCluyze\altaffilmark{28}, 
B.~C.~Lee\altaffilmark{33}, 
E.~Lindfors\altaffilmark{2}, 
J.~Lozano de Haro\altaffilmark{34}, 
J.~P.~Moore\altaffilmark{28}, 
M.~Mugrauer\altaffilmark{35}, 
R.~Naves Nogues\altaffilmark{15}, 
A.~W.~Neely\altaffilmark{36}, 
R.~H.~Nelson\altaffilmark{37}, 
W.~Ogloza\altaffilmark{4}, 
S.~Okano\altaffilmark{31}, 
J.~C.~Pandey\altaffilmark{27}, 
M.~Perri\altaffilmark{5,38}, 
P.~Pihajoki\altaffilmark{39}, 
G.~Poyner\altaffilmark{40}, 
J.~Provencal\altaffilmark{20}, 
T.~Pursimo\altaffilmark{41}, 
A.~Raj\altaffilmark{33}, 
D.~E.~Reichart\altaffilmark{28}, 
R.~Reinthal\altaffilmark{2}, 
S.~Sadegi\altaffilmark{2}, 
T.~Sakanoi\altaffilmark{31}, 
J.-L.~Salto Gonz{\'a}lez\altaffilmark{42},
Sameer\altaffilmark{12}, 
T.~Schweyer\altaffilmark{43}, 
%%V.~Simon\altaffilmark{29}, 
M.~Siwak\altaffilmark{4}, 
F.~C.~ Sold{\'a}n Alfaro\altaffilmark{44}, 
E.~Sonbas\altaffilmark{45}, 
I.~Steele\altaffilmark{11}, 
J.~T.~Stocke\altaffilmark{46},
J.~Strobl\altaffilmark{29}, 
L.~O.~Takalo\altaffilmark{2}, 
T.~Tomov\altaffilmark{47}, 
L.~Tremosa Espasa\altaffilmark{48}, 
J.~R.~Valdes\altaffilmark{19}, 
J.~Valero P{\'e}rez\altaffilmark{49}, 
F.~Verrecchia\altaffilmark{5,38}, 
J.~R.~Webb\altaffilmark{50}, 
M.~Yoneda\altaffilmark{51}, 
M.~Zejmo\altaffilmark{52},
W.~Zheng\altaffilmark{24},
J.~Telting\altaffilmark{41},
J.~Saario\altaffilmark{41},
T.~Reynolds\altaffilmark{41},
A.~Kvammen\altaffilmark{41},
E.~Gafton\altaffilmark{41}, 
R.~Karjalainen\altaffilmark{53},
J.~Harmanen\altaffilmark{2}
and P.~Blay\altaffilmark{54}
}
  
\altaffiltext{1} {Finnish Centre for Astronomy with ESO, University of Turku, Finland}
\altaffiltext{2} {Tuorla Observatory, Department of Physics and Astronomy, University of Turku, Finland}
\altaffiltext{3} {Astronomical Observatory, Jagiellonian University, ul. Orla 171, Cracow PL-30-244, Poland}
\altaffiltext{4} {Mt. Suhora Astronomical Observatory, Pedagogical University, ul. Podchorazych 2, PL30-084 Cracow, Poland} 
\altaffiltext{5} {Agenzia Spaziale Italiana (ASI) Science Data Center, I-00133 Roma, Italy}
\altaffiltext{6} {Istituto Nazionale di Fisica Nucleare, Sezione di Perugia, I-06123 Perugia, Italy}
\altaffiltext{7} {Department of Astronomy and Astrophysics, Tata Institute of Fundamental Research, Mumbai 400005, India}
\altaffiltext{8} {Astronomical Institute, Osaka Kyoiku University, 4-698 Asahigaoka, Kashiwara, Osaka 582-8582, Japan}
\altaffiltext{9} {Herschel Science Centre, ESAC, European Space Agency, 28691 Villanueva de la Ca{\~n}ada, Madrid, Spain}
\altaffiltext{10} {Department of Astrophysics, Astronomy and Mechanics, National \& Kapodistrian University of Athens, Zografos GR-15784, Athens, Greece}
\altaffiltext{11} {Astrophysics Research Institute, Liverpool John Moores University, IC2, Liverpool Science Park, Brownlow Hill, L3 5RF, UK}
\altaffiltext{12} {Physical Research Laboratory, Ahmedabad 380009, India}
\altaffiltext{13} {Department of Physics, Faculty of Arts and Sciences, Canakkale Onsekiz Mart University, TR-17100 Canakkale, Turkey; Astrophysics Research Center and Ulupinar Observatory, Canakkale Onsekiz Mart University, TR-17100, Canakkale, Turkey}
\altaffiltext{14} {5, Silver Lane, West Challow, Wantage, Oxon, OX12 9TX, UK}
\altaffiltext{15} {C/ Jaume Balmes No 24 08348 Cabrils, Barcelona, Spain}
\altaffiltext{16} {C/.Riera, 1, 1$^o$ 3$^a$ Barcelona, Spain}
\altaffiltext{17} {Carretera de Martos 28 primero Fuensanta, Jaen, Spain}
\altaffiltext{18} {Dark Sky Observatory, Dept. of Physics and Astronomy, Appalachian State University, Boone, NC 28608, USA}
\altaffiltext{19} {Instituto Nacional de Astrofisica, \'Optica y Electr\'onica, Apartado Postal 51-216, 72000 Puebla, M\'exico}
\altaffiltext{20} {University of Delaware, Department of Physics and Astronomy, Newark, DE, 19716, USA}
\altaffiltext{21} {Institute of Astronomy and NAO, Bulg. Acad. Sc., 72 Tsarigradsko Chaussee Blvd., 1784 Sofia, Bulgaria}

\altaffiltext{22} {Department of Astronomy and Astrophysics, Ataturk University, Erzurum, 25240, Turkey}
\altaffiltext{23} {Aritz Bidea No 8 4B (48100) Mungia Bizkaia, Spain}
\altaffiltext{24} {Department of Astronomy, University of California, Berkeley, CA 94720-3411, USA}
\altaffiltext{25} {Mu\~nas de Arriba La Vara, Vald{\'e}s (MPC J38) 33780 Vald\'es, Asturias -- Spain}
\altaffiltext{26} {C/ Concejo de Teverga 9, 1C 28053 Madrid, Spain}
\altaffiltext{27} {Aryabhatta Research Institute of Observational Sciences (ARIES), Nainital, 263002 India}
\altaffiltext{28} {University of North Carolina at Chapel Hill, Chapel Hill, North Carolina NC 27599, USA}
\altaffiltext{29} {Astronomical Institute, The Czech Academy of Sciences, 25165 Ond{\v r}ejov, Czech Republic; Czech Technical University in Prague, Faculty of Electrical Engineering, Prague, Czech Republic}
\altaffiltext{30} {16 Westminster Close Basingstoke Hampshire RG22 4PP, UK}
\altaffiltext{31} {Planetary Plasma and Atmospheric Research Center, Tohoku University, Sendai, Japan}
\altaffiltext{32} {Department of Physics and Astronomy and SARA Observatory, University of Alabama, Box 870324, Tuscaloosa, AL 35487, USA}
\altaffiltext{33} {Korea Astronomy and Space Science Institute, 776, Daedeokdae-Ro, Youseong-Gu, 305-348 Daejeon, Korea 
                   and Korea University of Science and Technology, Gajeong-Ro Yuseong-Gu, 305-333 Daejeon,Korea}
\altaffiltext{34} {Partida de Maitino, pol. 2 num. 163 (03206) Elche, Alicante, Spain}
\altaffiltext{35} {Astrophysikalisches Institut und Universit\"ats-Sternwarte, Schillerg\"a{\ss}chen 2-3, D-07745 Jena, Germany}
\altaffiltext{36} {NF/Observatory, Silver City, NM 88041, USA}
\altaffiltext{37} {1393 Garvin Street, Prince George, BC V2M 3Z1, Canada}
\altaffiltext{38} {INAF--Osservatorio Astronomico di Roma, via Frascati 33, I-00040 Monteporzio Catone, Italy}
\altaffiltext{39} {Department of Physics, University of Helsinki, P.O. Box 64, FI-00014 Helsinki, Finland}
\altaffiltext{40} {BAA Variable Star Section, 67 Ellerton Road, Kingstanding, Birmingham B44 0QE, UK}
\altaffiltext{41} {Nordic Optical Telescope, Apartado 474, E-38700 Santa Cruz de La Palma, Spain}
\altaffiltext{42} {Observatori Cal Maciarol m{\`o}dul 8. Masia Cal Maciarol, cam{\'i} de l'Observatori s/n 25691 {\`A}ger, Spain}	
\altaffiltext{43} {Max Planck Institute for Extraterrestrial Physics, Giessenbachstrasse, D-85748 Garching, Germany; Technische Universit\"at M\"unchen, Physik Department, James-Franck-Str., D-85748 Garching, Germany}
\altaffiltext{44} {C/Petrarca 6 1{$^a$} 41006 Sevilla, Spain}
\altaffiltext{45} {University of Adiyaman, Department of Physics, 02040 Adiyaman, Turkey}
\altaffiltext{46} {Center for Astrophysics and Space Astronomy, Department of Astrophysical and Planetary Sciences, Box 389, University of Colorado, Boulder, CO 80309, USA}
\altaffiltext{47} {Centre for Astronomy, Faculty of Physics, Astronomy and Informatics, Nicolaus Copernicus University, 
                   ul. Grudziadzka 5, 87-100 Torun, Poland}
\altaffiltext{48} {C/Cardenal Vidal i Barraquee No 3 43850 Cambrils, Tarragona, Spain}
\altaffiltext{49} {C/Matarrasa, 16 24411 Ponferrada, Le{\'o}n, Spain}
\altaffiltext{50} {Florida International University and SARA Observatory, University Park Campus, Miami, FL 33199, USA}
\altaffiltext{51} {Kiepenheuer-Institut fur Sonnenphysic, D-79104, Freiburg, Germany}
\altaffiltext{52} {Janusz Gil Institute of Astronomy, University of Zielona G{\'o}ra, Szafrana 2, PL-65-516 Zielona G{\'o}ra, Poland}
\altaffiltext{53} {Isaac Newton Group of Telescopes, Apartado  321, E-38700 Santa Cruz de La Palma, Spain} 
\altaffiltext{54} {IAC-NOT, C/Via Lactea, S/N, E38205, La Laguna, Spain}

\email{email: {\tt mvaltonen2001@yahoo.com}}
 
\begin{abstract}
OJ~287 is a quasi-periodic quasar with roughly 12 year optical cycles.  
It displays prominent outbursts which are predictable in a binary black 
hole model. The model predicted a major optical outburst in December 2015. 
We found that the outburst did occur within the expected time range, peaking 
on 2015 December 5 at magnitude 12.9 in the optical $R$-band. Based on 
{\it Swift}/XRT satellite measurements and optical polarization data, we find 
that it included a major thermal component. Its timing provides an accurate 
estimate for the spin of the primary black hole, $\chi = 0.313 \pm 0.01$.  
The present outburst also confirms the established general relativistic 
properties of the system such as the loss of orbital energy to gravitational 
radiation at the $2\%$ accuracy level and it opens up the possibility of testing 
the black hole no-hair theorem with a $10\%$ accuracy during the present decade.  
\end{abstract}


\section{Introduction}
OJ~287 is recognized as a quasar with roughly 12 year cycles in optical 
brightness, observed since 1890's \citep{sil88}.
Its light 
curve is definitely not periodic 
\citep{hud13}
but the deviations 
from periodicity are systematic and predictable in a model that contains a 
gravitational wave driven inspiralling spinning binary black hole system as 
its central engine 
\citep{val08b, val10a, byr15}.
The 
prediction for the 2015/6 observing season was that OJ~287 should have a major 
optical outburst in December 2015, brightest optical level in 30 years 
(see Figure 1 in \citealt{val11a}
for the future light curve), 
coinciding with the centenary of General Relativity. The exact timing of 
the optical outburst may be used to test predictions of the general 
relativistic  binary black hole model \citep{val97, val10b}. 
	
The quasi-periodic pattern of optical outbursts of OJ~287 was explained 
in 1995 by a model where a secondary black hole in a 12 year orbit 
impacts the accretion disk of the primary black hole at regular intervals 
\citep{leh96, sun97}.
Owing to the quasi-Keplerian 
nature of binary black hole orbits in general relativity, the impacts and 
their associated electromagnetic radiation events cannot occur in a strictly 
periodic manner \citep{dam88, mem04}. 
Attempts to use purely Newtonian orbit models, ignoring post-Newtonian 
corrections, have failed  \citep{val11c, val12a}.
\footnote{For a pictorial depiction of binary black hole 
orbits in general relativity, see the animations by S. Drasco at 
www.tapir.caltech.edu/$\sim$sdrasco/animations.}

However, it is indeed possible to find a unique mathematical description 
for the orbit in the post-Newtonian approximation to general relativity, 
provided a long enough record of past radiation outbursts is available. 
The solution is sensitive to the loss in gravitational binding energy caused by 
gravitational wave emission and the Lense-Thirring effect that forces the 
binary black hole orbital plane to precess, mainly due to the spin of the 
primary black hole \citep{bar75, dam88}.

\begin{figure}[ht]
\includegraphics[angle=0, width=1.07\columnwidth] {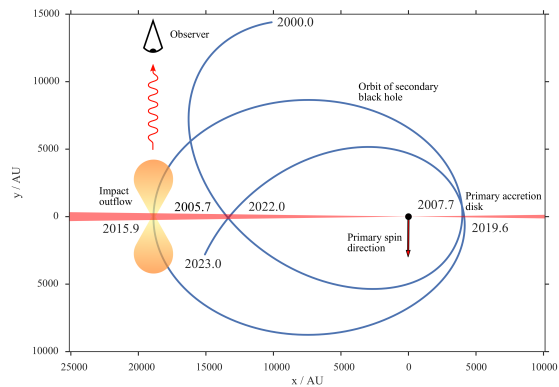}
\caption{The orbit of the secondary black hole in OJ~287 from year 2000 to 2023.
The present thermal outburst comes from the disk crossing in 2013
while the nonthermal flux arises from a jet, parallel to the primary spin 
axis. The next two thermal outbursts are due in 2019 and 2022, following 
the crossing of the secondary black hole through the accretion disk of the 
primary black hole.}
\label{fig1}
\end{figure}

An essential feature of the model is that the impact outbursts are generated 
by expanding bubbles of hot gas which have been shocked and pulled out of 
the accretion disk (see Figure \ref{fig1}). 
The process is astrophysically rather simple \citep{leh96, iva98, pih16} 
and the resulting radiation emanates from the vicinity of the impact site. Thus, 
these thermal radiation events are excellent markers for tracing the orbital 
motion of the secondary around the primary black hole. In contrast, the 
shocks in jets which also arise as a consequence of the influence of the 
secondary have a complicated route from cause to effect. These "tidal" 
outbursts \citep{sil88, sun97, val09, val11b} 
are also predictable in the binary model, 
but cannot be used to construct the orbit as accurately as by using the record 
of the thermal events. In other systems the regularly repeated events in a jet 
may be a more likely alternative than thermal events \citep{ack15}.

\section{OBSERVATIONS}
\label{obs}

In anticipation of the predicted outburst, we organized a multisite optical observing campaign 
aimed at getting photometric and polarimetric data on OJ~287.
Both professional astronomical observatories and amateur observers took 
part in obtaining photometric data from the very beginning of the 2015/2016 season.
%  Here changes suggested by Mark
%Some of the telescopes of the amateur astronomers who contributed to the current campaign 
%are listed by \citet{val08a}.
The telescopes of amateur astronomers are in the 0.20$-$0.30~m class \citep{val08a}. 

\begin{figure*}[ht]
\includegraphics[width=9.5cm,height=17.2cm,scale=1.0,angle=270] {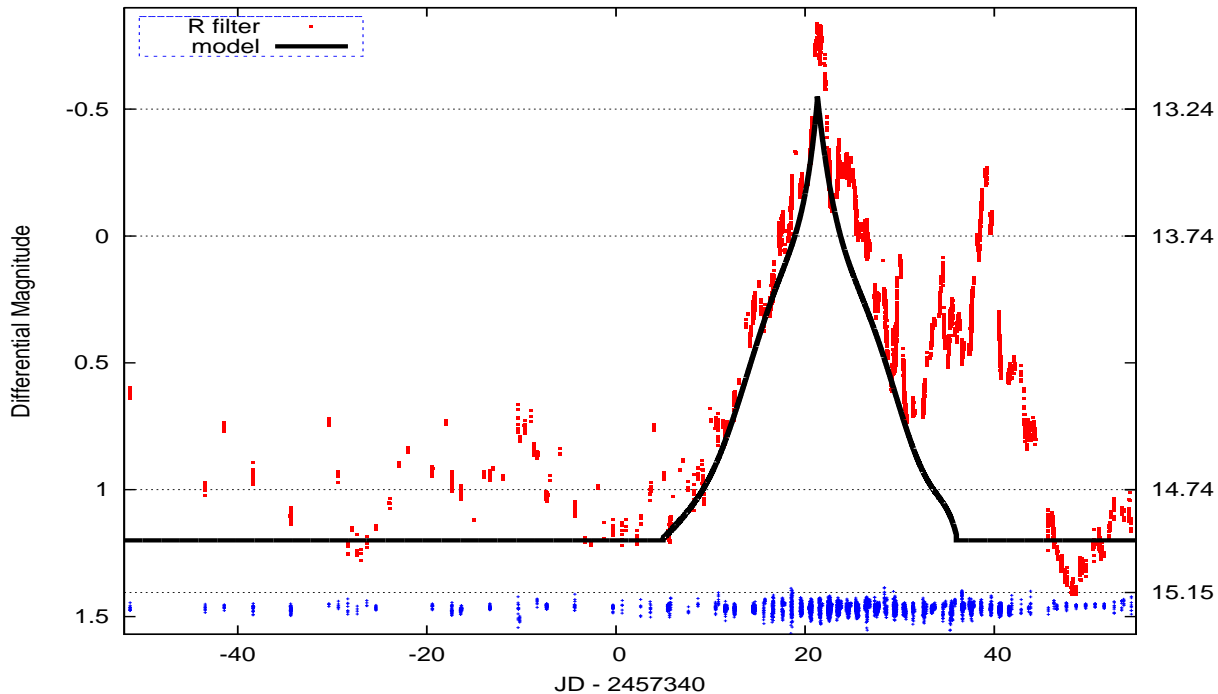}
\caption{
Optical photometry of OJ~287 from October to December 2015. The optical $R$-band
magnitude (squares) is given with respect to GSC~1400-222 comparison star.
%The symbols refer to observations made at different observatories. 
At the bottom (crosses) we show the differences between our comparison and the
check star (GSC~1400-444, shifted by 2.05 mag).
The theoretical line is explained 
in the text. }
\label{fig2}
\end{figure*}

Photometric observations were carried out by the following observatories: 
Tuorla Observatory in Finland, Mount Suhora Observatory of the Pedagogical
University and Astronomical Observatory of the Jagiellonian  University 
in Poland, University of Athens in Greece, 
Mount Abu Infrared Observatory in India, and Liverpool Telescope,
Kungliga Vetenskapliga Akademien Telescope, Nordic Optical Telescope (NOT), and William Herschel Telescope using ACAM instrument, in La Palma, Canary Islands, Spain 
(see \citet{pih13} for details).
Other telescopes participating were  0.41~m PROMPT5  
telescope in Chile \citep{rei05}, the 0.6~m SARA telescope at the Cerro Tololo InterAmerican Observatory, 
the 0.51~m reflector in Osaka Kyoiku University, Japan, 
the 0.25~m Cassegrain and 
0.9/0.6~m Schmidt telescopes of the University Observatory Jena, Germany \citep{mug10, mug16}, 
the 0.77~m Schmidt Camera of Tonantzintla in Mexico, the 0.60~m and 1.22~m reflectors of 
the Canakkale Onsekiz Mart University Observatory, the 0.60~m telescope of the University 
of Adiyaman and the 0.60~m telescope at the TUBITAK National Observatory, Turkey 
and the 0.50~m robotic telescope at the Ondrejov Observatory, Czech Republic.
In the continental US the photometric data were gathered with the 0.9~m 
SARA telescope at Kitt Peak, the 0.40~m telescope of Florida International 
University, the 0.76~m Katzman Automatic Imaging Telescope (KAIT) at the Lick Observatory \citep{fil01}, 
the 0.40~m University of Alabama campus telescope and the 
0.40~m Arizona State University campus telescope. 
OJ~287 was measured through the wide band $R$ filter in most sites. Only the KAIT data 
were taken without any filter and transformed into the $R$ band. 
We performed differential photometry on images calibrated for bias, dark and flatfield 
with the aperture method. We used GSC~1400-222 ($R=13.74$ mag) as the comparison star and 
GSC~1400-444 as the check star.

\begin{figure}[ht]
\includegraphics[angle=270, width=1.0\columnwidth] {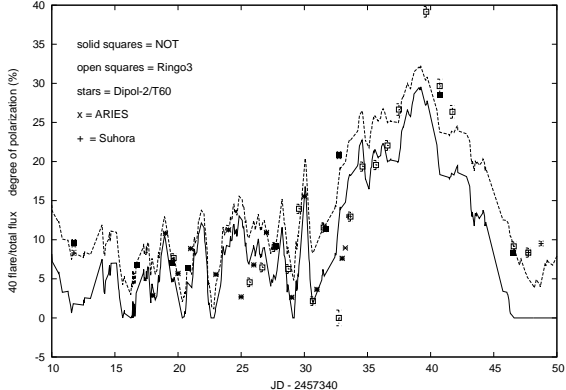}
\caption{The degree of polarization in the optical $R$-band. The curve represents
 the expected degree of polarization if the excess nonthermal component, 
above the line in Figure 2, is $40\%$ polarized and the rest of the radiation 
is unpolarized. The dashed line assumes that in addition the base level flux 
makes a $10\%$ contribution to the degree of polarization. Nightly median 
values are plotted for the Ringo3 observations.}
\label{fig3}
\end{figure}
 
Measurements with  the DIPOL-2 polarimeter \citep{pii14} 
installed on  the remotely controlled, 0.60~m telescope at the Haleakala observatory 
(Tohoku University) were carried out on 13 nights in the 
interval 2015 Nov 30 -- Dec 15 (UT dates are used throughout this paper).  
Simultaneous observations in three different passbands 
($B$, $V$, $R$) were made by using dichroic beam splitters to divide the light, 
which was then recorded by three CCDs. 
% Omit the next sentence
%With the 
%simultaneous observations in the different passbands, rapid variations in the 
%blazar polarization have no effect on the observed wavelength dependence. 
On each night, $32 \times 30$~s exposures of OJ~287 were obtained at different orientations 
($22.5^\circ$ steps) of the superachromatic half-wave retarder used as the 
polarization modulator. The fluxes of the target images on the CCD frames 
were extracted by using a circular aperture of $4''$--$6''$ radius.

Polarization and photometry observations of OJ~287 were taken on 20 nights (89 altogether) 
in the interval 2015 November 28 to 2015 December 31 with the RINGO3 
polarimeter \citep{arn12}
on the fully robotic and autonomous Liverpool 
Telescope on La Palma, Canary Islands  \citep{ste10}.
Simultaneous 
observations (120~s duration) in three passbands (blue, 3500--6400~\AA; 
green, 6500--7600~\AA; and red, 7700--10000~\AA) were taken using the rapidly 
rotating (once per 4 seconds) polaroid which modulates the incoming beam of 
light in 8 rotor positions, and for the photometry the 8 frames are stacked. 
The beam is simultaneously split by 2 dichroic mirrors into three low-noise electron multiplying
CCD cameras. The fluxes of the OJ~287 images on the EMCCD 
frames were extracted using circular apertures with radii of $4''$. 
By combining the flux from the 8 rotor positions using equations from \citet{cla02}
the linear Stokes parameters were measured and used 
to calculate the degree and angle of polarization. The data were corrected 
for the effects of instrumental polarization and depolarization by observation 
of standard stars from \citet{sch92}.

The polarimetric observations at the 2.5~m NOT telescope (Principal Investigator 
K.N.) were carried out in the manner described in \citet{val09}.
Polarization observations with the 0.60~m  Cassegrain telescope at Mount Suhora Observatory were carried out during 7 nights  using four polarimetric filters 
transmitting  light of the polarization planes $0^\circ$, $45^\circ$, $90^\circ$, and $135^\circ$. 
%One series of measurements in all four filters can be used to determine the degree 
%of polarization and the position angle. 
At least 7 full series of measurements in all four filters were performed 
each night with exposure times between 30~s and 90~s, depending on 
weather conditions and brightness of the target. 
Fluxes were extracted by using the IRAF "apphot" package.  Finally, the degree 
of polarization and position angles were obtained by the n-polarizers 
method proposed by \citet{spa99}.

\begin{figure}[]
%\centering
%\begin{minipage}{1.0\columnwidth} 
\includegraphics[angle=0,width=7.7cm,height=12.0cm,scale=1.0] {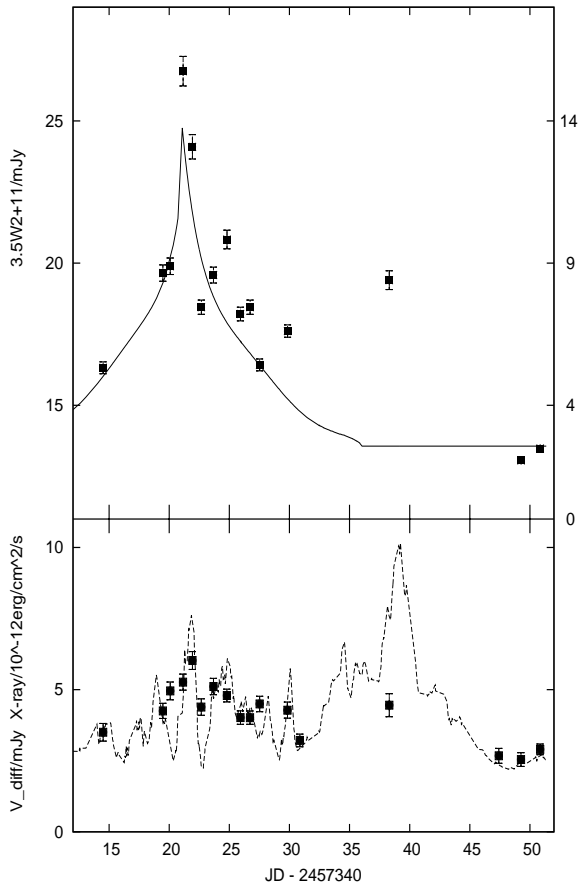}

\caption{
Bottom panel: A comparison of X-ray observations by \textit{Swift}/XRT in 
the 0.3--2 keV  energy band in ergs cm$^{-2}$ s$^{-1}$ (squares with errorbars) with the excess 
("jet") emission above the line of Figure \ref{fig2} (dashed line). Top panel: \textit{Swift}/UVOT observations in the ultraviolet (UV) {\it UVW2} band (central wavelength 1928~\AA) shown as squares with errorbars. The model line shown in Figure \ref{fig2} has been shifted to the {\it UVW2} band by using 
the spectral index of 1.35.
}
%\end{minipage}
\label{fig45}
\end{figure} 

Two polarimetric observations were obtained using the ARIES Imaging Polarimeter (AIMPOL; 
\citealt{rau04}), 
mounted on the 1.04~m Sampurnanand Telescope \citep{sin75} 
at Nainital, India, coupled with a TK $1024 \times 1024$ pixels CCD camera.

A time-domain program for the \textit{Swift} satellite dedicated to OJ~287
was performed (Principal Investigator S.C.) in parallel to these optical
multisite
observations, using \textit{Swift}/UVOT filters {\it UVW1}, {\it UVM2} and {\it UVW2} and {\it Swift}/XRT (0.3--10 keV band).
Here we report primarily the results from the ground-based optical
telescope $R$-band and \textit{Swift}/UVOT~{\it UVW2} band 
where the results were more complete than in other channels, in addition to 
X-rays. 
\textit{Swift}/XRT data were taken in photon-counting mode for a total exposure of about 
20~ksec divided into daily observations. Each single X-ray spectrum (0.3--10 keV) 
can be fit by an absorbed single (or broken) power-law model, with an H~I column 
density consistent with the Galactic one in the direction of the source 
($n_H = 2.56 \times 10^{20}$ cm$^{-2}$, \citealt{kal05}).
The X-ray spectra have photon indexes between about 1.4 and 1.9. The preliminary 
corresponding unabsorbed (0.3--2.0 keV) integral daily fluxes are reported here, 
together with 
simultaneous dereddened UVOT flux-density values obtained with the 3 UV filters.

%\begin{figure}[ht] 
%\epsscale{1.15} 
%\plotone{Letter_figure_4.eps}
%\includegraphics[angle=270, width=1.0\columnwidth] {Letter_figure_4.eps}
%\caption{A comparison of X-ray observations by \textit{Swift} XRT in 
%the 0.3-2 keV  energy band in ergs cm$^{-2}$ sec$^{-1}$ with the excess 
%('jet') emission above the line of Figure \ref{fig2}. 
%Compared with the optical outburst, the X-ray flare was rather modest, 
%consistent with the view that the thermal component 
%makes no detectable contribution in X-rays.}
%\label{fig4}
%\end{figure} 

%\begin{figure}[ht] 
%\epsscale{1.15} 
%\includegraphics[angle=270, width=1.0\columnwidth] {Letter_figure_5.eps}
%\caption{\textit{Swift} UVOT observations in the UV W2 band (central 
%wavelenght 1928 \AA) shown as points with errorbars. 
%The broken line below represents optical light curve from Figure \ref{fig2}, 
%while the smooth dashed line below it is the corresponding line in Figure 
%\ref{fig2}.
%This line has been shifted to the W2 band by using the spectral index 
%of 1.35.}
%\label{fig5}
%\end{figure} 

After starting intensive optical ground-based photometric monitoring of OJ~287 in September 2015,
a series of frames (about 10 images per night) were taken to measure 
the brightness of the target every clear night. 
Nightly means were calculated and posted on the campaign's web page. 
After November 14, a steady rise of the object flux was noticed, and by 
November 25 it was apparent that it may develop into a major outburst, 
in the category observed only twice in 12~yr. We extended our
observations by making them as long as possible each night, measuring also 
colors at some sites.  
The source kept brightening very rapidly until it was brightest it has been in 30~yr.  
After the December 5 maximum the source declined in stages, until it 
arrived at its pre-outburst level on December 30 (see Figure \ref{fig2}).

The major outbursts in OJ~287 are recognized by a rapid rise to a narrow 
peak and then a slower decline with multiple smaller flares. The general shape of the curve 
in Figure \ref{fig2} is based on a model of a uniform expanding sphere of plasma \citep{pih16}. However, as seen generally in observations {\it e.g.} in 
1983 \citep{smi87}, there is an initial, slowly rising part in the light curve with an additional peak at the maximum. This ``standard light curve'' \citep{val11a} is used here even though detailed theoretical models do not exist.

The information on the nature of radiation at different 
stages of the outburst has been limited up to now. In 2007 a good coverage 
of the outburst was achieved in polarization; it showed that the major 
component of the outburst was unpolarized, superposed on a lower level 
of polarized synchrotron emission \citep{val08b}.
In 1983 the degree 
of polarization decreased close to zero at the high point of the light curve 
\citep{smi87}.
Therefore we have reasons to believe that an underlying 
unpolarized component, like the curve in Figure \ref{fig2}, 
also exists in the 2015 outburst, in addition to the usual polarized flares.

Figure \ref{fig3} shows the evolution of the degree of polarization at 
different stages of the 2015 outburst. We superpose on the data the expected 
degree of polarization, by using the theoretical line in Figure \ref{fig2} 
to separate the thermal and nonthermal components of the outburst. 
The solid line gives the ratio of the excess radiation above the theoretical
line to the total flux, multiplied by 40. This is what one might expect 
if the radiation below the theoretical line is unpolarized, as thermal 
bremsstrahlung should be, and superposed on it we have synchrotron flares 
with $40\%$ polarization. The second line assumes an additional $10\%$ contribution to the degree of polarization from the base level flux. This simple concept seems to work reasonably well. 
If our separation of the bremsstrahlung from synchrotron flares in the model 
is correct, then the X-ray emission, coming entirely from the jet, should 
follow the optical excess emission. The optical excess emission is defined 
as the total optical flux minus the bremsstrahlung flux, the latter separated 
from the total flux according to the line in Figure \ref{fig2}. 

Figure \ref{fig45} shows that this is indeed the case. The X-ray flare is rather modest, much smaller than the optical outburst overall, but correlates very well with the excess flare emissions. The flares arising 
at this time are not different from flares observed during the campaigns of 
the previous twelve months (Edelson et al. 2015). There the X-ray flux was 
$(4.0 \pm 1) \times 10^{-12}$ ergs cm$^{-2}$ s$^{-1}$, while during our campaign it has 
been $(4.4 \pm 1) \times 10^{-12}$ ergs cm$^{-2}$ sec$^{-1}$, only slightly enhanced.

The UV emission has followed the optical emission rather well in previous 
campaigns, using a spectral index of 1.35 between the two wavelength ranges 
(based on data from the \citealt{ede15} campaign) 
of \textit{Swift}/UVOT. The figure shows that the 
new line in the {\it UVW2} channel follows the data rather well. Above the thermal 
components there are the same nonthermal flares that are seen in optical. 
The \textit{Swift} {\it UVW1} and {\it UVM2} band results are entirely consistent 
with Figure \ref{fig45}. A more careful study is required to determine the 
temperature of the bremsstrahlung component at this time \citep{val12b}.
 
\section{DISCUSSION}
\label{discussion}

The timing signals are extracted from the optical light curve by identifying 
the start of the outburst. From Figure \ref{fig2} it appears that the outburst 
began on JD~2,457,342.5$\pm$2.5 which corresponds to year $2015.874 \pm 0.007$. 
Using the previously calculated correlation with the spin \citep{val11a},
we get for the Kerr parameter of the primary black 
hole $\chi = 0.313 \pm 0.01$ (2 $\sigma$). We have checked with orbit solutions, making use of 
this new timing, that the possible $\chi$ values range between 0.304 and 0.322. This is a considerable 
improvement with respect to the previous value $\chi = 0.28 \pm 0.08$ 
\citep{val10a}.

For a comparison with black hole spin determinations by X-ray spectroscopy, 
see  \citet{rey08} and \citet{rey14}.
These are based on 
determining the innermost stable orbit of the accretion disks in Seyfert nuclei 
or low-redshift quasistellar objects (QSOs) in the radio-quiet realm, or in X-ray binaries. 
Some of the spins are comparable to the spin of OJ~287, others are close to 
the maximal value of unity, while the recently observed merger of two black holes produced a spin value of 0.67 \citep{abb16}. In contrast to the X-ray spectroscopy method, 
in this blazar we are not dependent on understanding the physics of accretion disks 
close to the innermost stable orbit; in this sense the orbital torque method 
is complementary to X-ray spectroscopy.

The present outburst timing firmly confirms the correctness of the binary 
black hole central engine model for OJ~287 within its specified parameter 
ranges, namely primary mass $(1.83 \pm 0.01) \times 10^{10}$~M$_\odot$, secondary 
mass $(1.5 \pm 0.1) \times 10^8$~M$_\odot$ and orbital eccentricity (as defined 
by using the apocentre/pericentre ratio) $0.700 \pm 0.001$.

The present $\chi$ estimate opens up the possibility of measuring the 
dimensionless quadrupole moment of the primary black hole ($q_2$) at the 
$10\%$ level during the next thermal outburst, predicted to happen in 
July 2019 (see Figure \ref{fig1}). This should allow one to test the 
black hole no-hair theorem by verifying the relation
 $q_2 = -\chi^2$ at that level \citep{car70, tho85}.
However, observing the predicted July 2019   
thermal outburst from the Earth will be difficult owing to the proximity of 
OJ~287 to the Sun at that time.

Additionally, as demonstrated earlier \citep{val10b, val11a},
the occurrence of the outburst within the expected 
time window confirms the loss of energy by gravitational radiation within 
$2\%$ of the prediction by general relativity and is consistent with 
the no-hair theorem of black holes within an accuracy of $30\%$.
The energy loss from impacts on the accretion disk is four orders of magnitude 
smaller than the energy loss through gravitational radiation and thus plays no 
role in the binary model.

Finally we note that an exceptionally large amount of gas has been 
pulled away from the primary disk during this impact which occurred close 
to the apocentre of the binary orbit (\citet{pih13}, see Figure \ref{fig1}). 
This gas is expected to feed the two black holes for 
some time to come, and keep OJ~287 active with flares.
 
The highly polarized ($39\%$ polarization) flare near JD~2,457,380 is 
interesting; its degree of polarization is the highest ever 
measured in OJ~287. The previous record was $36\%$ polarization measured 
in the secondary peak of the 1984 major event \citep{smi87}.
This suggests that the secondary flare is closely connected with the first, 
unpolarized outburst. One possibility is the activation of the jet of 
the secondary black hole at these times. The secondary black hole is in 
the vicinity of the expanding cloud of plasma and will definitely accrete 
a major part of it --- that is, the part which is expanding to its direction. 
It will be interesting to search for other evidence to associate the 
secondary flare with the secondary black hole. 

In summary, we have shown that the outburst in OJ~287 in November--December 
2015 agrees with the binary black hole model, both with regard to the timing 
and the expected brightness as well as a major outburst component being thermal. The fact that such thermal outbursts are excellent 
trackers of the secondary black hole orbit allowed us to estimate the spin 
value of the primary more narrowly than before, $\chi = 0.313 \pm 0.01$. 
This November--December outburst firmly confirms the presence of an inspiraling massive 
black hole binary in OJ~287.
It therefore  makes a fitting contribution to
general relativity centenary celebrations of 2015/2016, and adds to the excitement over the first direct observation of a transient gravitational-wave signal \citep{abb16}.

\acknowledgments 
The \textit{Swift} gamma-ray burst explorer mission is part of NASA's medium 
explorer (MIDEX) program, led by NASA with participation of Italy and the UK. 
We would like to thank the \textit{Swift} team for making these observations 
possible, in particular D. Malesani as the \textit{Swift}  Observatory Duty 
Scientist. The authors acknowledge support by the following grants: NCN 
~2013/09/B/ST9/00599 ~(S.Z.) and ~DEC-2011/03/D/ST9/00656 ~(M.Z.), ~GACR~13-33324S ~(R.H.)
and Academy of Finland grant no. 1274931 (P.P.). A.V.F. and W.-K.Z. are greatful for financial assistance from NSF grant AST-1211916, NASA grant NNX12AF12G, the TABASGO Foundation, and the Christopher R. Redlich Fund. 
We extend our thanks to TUBITAK for partial support in using the T60 telescope with 
project number 10CT60-76. We are grateful to S. Masda, D. Wagner, T. Zehe, J. Greif, 
and H. Gilbert for their help in carring out some of the observations 
taken at the University Obervatory Jena, which is operated by the Astrophysical 
Institute of the Friedrich-Schiller-University.

KAIT and its ongoing operation were made possible by donations from Sun Microsystems, Inc., the Hewlett-Packard Company, AutoScope Corporation, Lick Observatory, the NSF, the University of California, the Sylvia and Jim Katzman Foundation, and the TABASGO Foundation. Research at Lick Observatory is partially supported by a generous gift from Google.

\end{document}